\documentclass[12pt]{article}
\usepackage{graphicx}
\usepackage{amsmath}
\usepackage{subfigure}
\usepackage{graphicx}
\usepackage{amssymb}
\usepackage{cite}
\usepackage{color}

\begin{document}
\begin{center}
{\Large\bf   Viability of Variable Generalised Chaplygin gas - a thermodynamical approach }\\[8 mm]
D. Panigrahi\footnote{ Sree Chaitanya College, Habra 743268, India
\emph{and also} Relativity and Cosmology Research Centre, Jadavpur
University, Kolkata - 700032, India , e-mail:
dibyendupanigrahi@yahoo.co.in }
and S. Chatterjee\footnote{Relativity and Cosmology Research Centre,
Jadavpur University, Kolkata - 700032, India, e-mail : chat\_sujit1@yahoo.com}
\\[10mm]

\end{center}
\begin{abstract}
The viability of the variable generalised Chaplygin gas (VGCG)
model is analysed from the standpoint of its thermodynamical
stability criteria with the help of an equation of state, $P = -
\frac{B}{\rho^{\alpha} }$, where $B = B_{0}V^{-\frac{n}{3}}$. Here
$B_{0}$ is assumed to be a positive universal constant,  $n$ is a
constant parameter and $V$ is the volume of the cosmic fluid. We
get the interesting result  that if the well-known stability
conditions of a fluid is adhered to,  the  values of  $n$ are
constrained  to be negative definite to make  $ \left(
\frac{\partial P}{\partial
 V}\right)_{S} <0$  \& $ \left( \frac{\partial P}{\partial
 V}\right)_{T} <0$ throughout the evolution.
 Moreover the positivity of thermal capacity at constant volume $c_{V}$ as
 also the validity of the third law of thermodynamics are ensured
 in this case.
 For the particular case  $n = 0$ the effective equation of state reduces
 to $\Lambda$CDM model in the late stage of the universe
 while for $n <0$ it mimics a  phantom-like cosmology which is in broad agreement with the
 present SNe Ia constraints like VGCG model. The  thermal
 equation of state is discussed and the EoS parameter is found to be an explicit
 function of temperature only.
 Further for large volume the thermal equation of
state parameter  is identical with the caloric equation of state
parameter when $ T \rightarrow 0$. It may also be mentioned that
like Santos et al our model does not admit of any critical points.
We also observe
 that although the earlier model of Lu explains many of the current observational
  findings of different probes it fails to explain the crucial tests of thermodynamical stability.
\end{abstract}

   ~~~~KEYWORDS : cosmology; chaplygin gas; thermodynamics\\
   \vspace{-0.5cm }


\section*{1. Introduction}

The discovery of  cosmic acceleration of the universe ~\cite{reis,
nop, aman}  has added a new challenge for fundamental theories of
physics and cosmology. NASA's  observations ~\cite{spe1,spe2} show
that the kind of matter of which stars and galaxies are made forms
less than $5 \%$ of the universe's total mass. Several independent
observations indicate that about $73 \%$ of the total energy
density of the universe is in the form of a mysterious dark energy
or gravitationally repulsive energy, and about $22 \%$ is in the
form of non-baryonic cold dark matter particles which clump
gravitationally, but which have never been directly detected.

The late acceleration of the universe is often attributed to the
presence of one of the most weird and mysterious stuffs termed
\emph{dark energy} in the cosmic fluid. But understanding the
nature of this dark energy is definitely one of the most
challenging theoretical problems facing present-day cosmology and
people embark on looking out for different plausible alternatives
with missionary zeal. For sparing the readers going through the
repetition of those arguments and its counterparts (for excellent
reviews in this field one is referred to for instance ~\cite{neu,
sah} )  we skip the details and mention that different variants of
Chaplygin gas are also serious contenders for the same.
Incidentally many workers in this field strongly feel that it is
not enough to keep on generalizing the original Chaplygin gas by
introducing new parameters to  the theory without much physical
basis   and then manipulate the arbitrary parameters at will to
explain  the observational results coming from different cosmic
probes but one should also concentrate on basic physics involved.
Returning to the multiplicity of arbitrary constants for greater
maneuverability one should also check  that the different forms of
the Chaplygin gas should behave as a closed thermodynamical
system. In this context, among many other things the question of
its stability is of utmost importance and we are primarily
motivated by the consideration if the well known stability
criteria may put some stringent conditions on the values of the
parameters of the system. In the  process we find that some of the
claims made by past workers are clearly ruled out when posited
against the stability criteria. With this in mind we have in the
past worked on the stability problem in some of  our works
~\cite{dp2,dp4} and this one  also deals with this specific
problem in the case of variable modified Chaplygin gas. A
Chaplygin type of gas cosmology~\cite{kam, ben} is one of the
plausible explanations of recent phenomena, which  is a new matter
field to simulate dark energy. This type of equation of state(EOS)
is not applicable in the case of primordial universe~ \cite{
ben,dp}. Such equation of state leads to a component which behaves
as dust at early stage and as cosmological constant ($\Lambda$) at
later stage. The form of the equation of state (EoS) of matter is
the following,

\begin{equation}\label{eq:eq1}
P =  - \frac{B}{\rho^{\alpha} }.
\end{equation}

Here $P$ corresponds to the pressure of the fluid and $\rho$ is
the energy density of that fluid and $B$ is a constant. Recently a
variable generalised Chaplygin gas (VGCG) model is proposed and
constrained using Union supernovae sample and Barion acoustic
 oscillation (BAO) ~\cite{lu1} assuming the
 $B$ to depend on scale factor of FRW metric. Now we have taken the
above relation as  $B = B_{0}V^{-\frac{n}{3}}$ where $n$ is an
arbitrary constant and $B_{0}$ an absolute constant.  $V$ is the
volume of the fluid. For $n = 0$ the VGCG equation of state
reduces to the generalised Chaplygin gas equation of
state~\cite{ben}. In the above mentioned work of Lu they got $n =
0.75$ and $\alpha = 1.53$.
 Later Lu et al ~\cite{lu2} studied the same case with $\alpha =1$ (VCG) with the help of
  supernovae union  and BAO data. The best fit values of model parameters are
   found to be $n = 1.30^{+0.46}_{-0.07} (1 \sigma)~  ^{+0.74}_{-0.81} (2 \sigma)$.
   Interestingly if SNe Ia Union data is only considered, they obtained
    $n= 0.13^{+1.42}_{-1.94}(1 \sigma) ~^{+2.14}_{-3.55}(2 \sigma)$. Again Sethi
\emph{et al}~\cite{set} showed that for $\alpha = 1$ the best fit
values of  $n$ lie in the interval $(- 1.3, 2.6)$ [\emph{WMAP 1st
Peak + SNe Ia(3$\sigma$})] and $(- 0.2, 2.8)$ [\emph{WMAP 3rd Peak
+ SNe Ia(3$\sigma$})].
 So from what has been discussed above we see that the value of $n$
 may be both positive or negative. Positive value of $n$ gives
 a quiescence type of evolution and big rip is avoided while the negative
  value of $n$  favors a
phantom-like Chaplygin gas model which allows for the possibility
of the dark energy density increasing with time. Relevant to
mention that recently there are some indications that a strongly
negative equation of state, $w \leq - 1$, may give a better fit
~\cite{guo1, guo2, alen} with observations.

Literature abounds with work where different forms of Chaplygin gas were taken as
dark energy model and tried to fit the different findings of the cosmological probes.
 But it is not enough to be able to match the observational data. But the
 thermodynamical viability of the fluid should also be seriously explored.
 In this context stability of the fluid is crucially important. Previously
 Santos et al ~\cite{san1} have studied and showed that the generalised Chaplygin gas ( GCG)
 model is thermodynamically stable.  The present work is motivated by the
  consideration if the variable generalised Chaplygin gas (VGCG) is also thermodynamically
  viable. In the process we find the thermodynamical stability criteria are satisfied
 subject to the condition that $n$ should be negative definite which apparently
contradicts Lu's ~\cite{lu1} conclusion for VGCG model. But
$\alpha =1$, the negative value of $n$ lies within the range
obtained by Sethi et al ~\cite{set} and also agreed with another
work of Lu et al ~\cite{lu2} where they have considered SN Ia Union data only.\\

  Following usual thermodynamical procedure we have derived expressions of different thermal
   quantities as functions of temperature and volume. We also find that the third law of
   thermodynamics is satisfied in the case of VGCG. In the cosmological context, we have also
    found that VGCG is also amenable to a unified picture of dark matter and energy which cools
    down as the universe expands. Finally, the paper ends with a brief discussion.

\section*{2. Energy equation}
One may take
\begin{equation}\label{eq:eq2}
\rho = \frac{U}{V},
\end{equation}
where $U$ and $V$ are the internal energy and volume filled by the
fluid respectively.

Now, we try to find out the energy  $U$ and pressure $P$ of Variable
Generalised Chaplygin gas (VGCG ) as a function of its entropy $S$ and volume
$V$. From  thermodynamics, one has the following
relationship
\begin{equation} \label{eq:eq3}
\left(\frac{\partial U }{\partial V}\right)_{S}= - P.
\end{equation}
With the help of  equations \eqref{eq:eq1}, \eqref{eq:eq2} and \eqref{eq:eq3}

\begin{equation}\label{eq:eq4}
\left(\frac{\partial U }{\partial V}\right)_{S}=
B_{0}V^{-\frac{n}{3}} \frac{V^{\alpha}}{U^{\alpha}},
\end{equation}
integrating,

\begin{equation} \label{eq:eq5}
U = \left[ \frac{3B_{0}(1+\alpha)V^{\frac{3(1+\alpha)-n}{3}}}
{3(1+\alpha)-n} + c
\right]^{\frac{1}{1+\alpha}}.
\end{equation}
where the parameter $c$ is the integration function, which may   either be
  a function of entropy $S$ only or a
universal constant. The equation \eqref{eq:eq5} may be recast as

\begin{equation}\label{eq:eq6}
U  = \left[\frac{B_{0}(1+\alpha)V^{-\frac{n}{3}}}{N}\right]^{\frac{1}{1+\alpha}}
V \left[1+
\left(\frac{\epsilon}{V}\right)^N\right]^{\frac{1}{1+\alpha}},
\end{equation}
where $N = \frac{3(1+\alpha)-n}{3}  > 1$ for
$(1+\alpha)> \frac{n}{3}$  for real $U$ and

\begin{equation}\label{eq:eq7}
\epsilon =
\left[\frac{3(1+\alpha)-n}{3B_{0}(1+\alpha)}~c\right]^{\frac{1}{N}}
= \left[\frac{Nc}{B_{0}(1+\alpha)}\right]^{\frac{1}{N}},
\end{equation}
which has  dimension of volume. Now the energy density $\rho$ of

the VGCG comes out to be

\begin{equation}\label{eq:eq8}
\rho = \left[\frac{B_{0}(1+\alpha)V^{-\frac{n}{3}}}{N}\right]^{\frac{1}{1+\alpha}}
 \left[1+
\left(\frac{\epsilon}{V}\right)^N\right]^{\frac{1}{1+\alpha}} .
\end{equation}
In what follows we shall try to  obtain the expressions of relevant physical
 quantities and investigate their behaviour. It is to be mentioned that we get
  exactly similar expression for density from energy conservation  equation  with
  the scale factor given by $a^3(t) = V $.\\

\section*{3. Thermodynamical Behaviour}

Using equations  \eqref{eq:eq1} and  \eqref{eq:eq8} the pressure $P$ of the VGCG
 may also be determined as a function of
entropy $S$ and volume $V$  in the following form

\begin{equation}\label{eq:eq9}
 P = - \left(B_{0} V^{-\frac{n}{3}} \right)^{\frac{1}{1+\alpha}}
 \left[\frac{N}{(1+ \alpha) \left \{1 + \left(\frac{\epsilon}{V} \right)^{N}\right\}}
  \right]^{\frac{\alpha}{1+\alpha}}.
\end{equation}

\begin{figure}[ht]
\begin{center}
   \includegraphics[width=8cm]{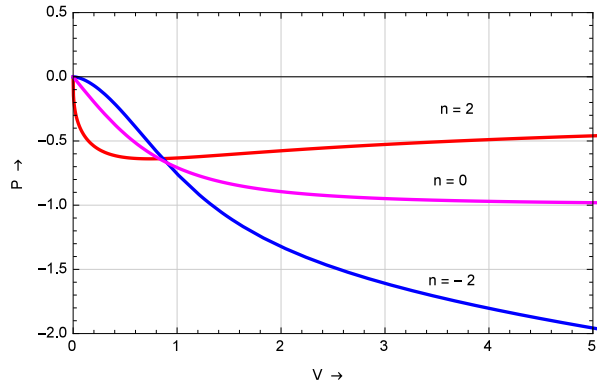}
     \caption{
  \small\emph{
  The variation of $p$ and $V$  for different
  values of $n$ with $ B_{0} =1$, $\alpha = 1$ \& $c =1$.
     }\label{1}
    }
\end{center}
\end{figure}

The equation \eqref{eq:eq9} gives a very general expression of
pressure where we see that $P$ is always negative. For $n=0$  the equation \eqref{eq:eq9}   becomes
\begin{equation}\label{eq:eq91}
 P = - \frac{(B_{0})^{\frac{1}{1+\alpha}}}{\left \{1+\left(\frac{c}{B_0 V}\right)^{1+\alpha}
  \right \}^{\frac{\alpha}{1+\alpha} }},
\end{equation}
which reduces to an earlier work of Santos et al ~\cite{san1} for the  generalised Chaplygin gas (GCG)
model. Again for  $ \alpha = 1$ in equation \eqref{eq:eq9}, we get
 $ P = - \left[ \frac{N B_0 V^{-\frac{n}{3}}}{2 \left\{1 + \left(\frac{ \epsilon}{V}
 \right)^N\right\}}\right]^{\frac{1}{2}}$ which is identical with our previous
 work ~\cite{dp2} when we consider the Variable Chaplygin gas (VCG) model.
\\
Now using  expressions   \eqref{eq:eq8} and  \eqref{eq:eq9} we get the \emph{effective}
equation of state parameter
\begin{equation}\label{eq:eq11}
\mathcal{W }= \frac{P}{\rho} =  \left\{ -1 +\frac{n}{ 3(1+ \alpha)}
\right\} \frac{1}{1 + \left(\frac{\epsilon}{V}\right)^N } .
\end{equation}

\begin{figure}[ht]
\begin{center}
   \includegraphics[width=8cm]{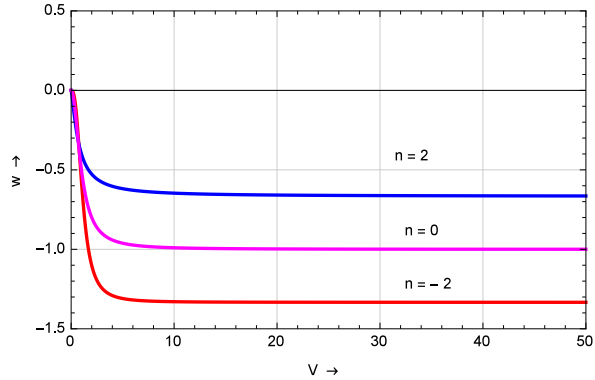}
     \caption{
  \small\emph{
   $w$ $\sim$ $V$ graph  for different
  values of $n$ with $B_{0} =1$, $\alpha = 1$ \& $c =1$.
     }\label{2}
    }
\end{center}
\end{figure}

Due to its complexity, we can not arrive at definite conclusions from equation \eqref{eq:eq11},
 which forces us to look forward to its extreme cases:
 For small volume, $V \ll \epsilon$, the equation  \eqref{eq:eq11} gives

\begin{equation}\label{eq:eq12a}
 P \sim 0 .
 \end{equation}
 Interestingly we see that for this dust dominated universe, the $n$ has essentially no influence on
 the pressure $P$. Again for large volume, $V \gg \epsilon$,  the equation \eqref{eq:eq11} becomes

\begin{equation}\label{eq:eq13}
\mathcal{W }\approx -1 + \frac{n}{3(1+\alpha)}.
\end{equation}
since   $n < 3(1+ \alpha)$. Three possibilities exist for  $\mathcal{W }$ depending on the
  signature of $n$ as
  (i) $n > 0$, $ \mathcal{W } > -1$, which points to a quiescence type of evolution
  and big rip is avoided in this case, (ii) $n = 0$, $\mathcal{W } = -1$. It represents
    $\Lambda$CDM and
  (iii) $n <0$, $\mathcal{W } < -1$ gives a phantom like universe.
  This is compatible with recent observational results ~\cite{guo1,guo2,alen}.
  Influence of $n$ is prominent in
 this case.\\

 Now  we calculate the deceleration parameter of the VGCG fluid from equation \eqref{eq:eq11} as
\begin{equation}\label{eq:eq14}
q = \frac{1}{2} + \frac{3}{2} \frac{P}{\rho} = \frac{1}{2} -
\frac{3}{2} \left(\frac{N}{1+ \alpha} \right) \frac{1}{1
+ \left(\frac{\epsilon}{V}\right)^N } ,
\end{equation}

\begin{figure}[ht]
\begin{center}
   \includegraphics[width=8cm]{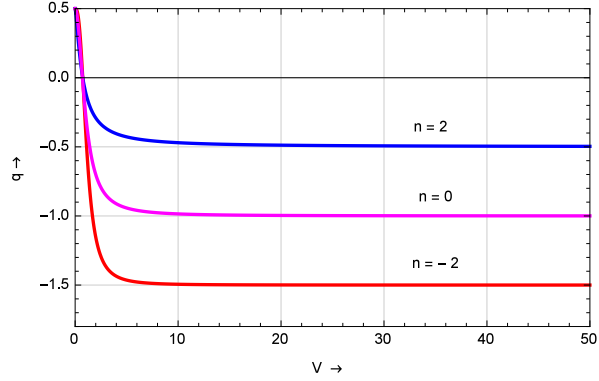}
     \caption{
  \small\emph{
  The variation of $q$ and $V$  for different
  values of $n$ with $ B_{0} =1$, $\alpha = 1$ \& $c =1$.
     }\label{3}
    }
\end{center}
\end{figure}
As the equation \eqref{eq:eq14} is very involved in nature we again look for extremal
 cases as before.
 For $V \ll \epsilon$, we get

\begin{equation}\label{eq:eq15}
q\sim \frac{1}{2}  ,
\end{equation}
\emph{i.e.},  universe decelerates for small $V$. Alternatively  for $V \gg \epsilon$,
the equation \eqref{eq:eq14} reduces to

\begin{equation}\label{eq:eq16}
q\approx -1 + \frac{n}{2(1+\alpha)} .
\end{equation}
To sum up  when volume is very small there is no
influence  of $n$  on $q$, here $q >0$, universe decelerates. But
for large volume,  $q <0 $  and  this  depends on the value of
$n$ also. For \emph{flip} in deceleration parameter  the flip volume
($V_{f}$) becomes

\begin{equation}\label{eq:eq17}
V_{f} = \epsilon \left[\frac{(1+
\alpha)}{2(1+\alpha)-n}\right]^{\frac{1}{N}}.
\end{equation}
The above equation dictates that $n < 2(1+ \alpha)$, which interestingly does not violate
our previous restriction on $n$. From equation \eqref{eq:eq17} it follows  that for
 $V_{f}$ to have real value $n < 2(1+
\alpha)$, otherwise \emph{flip} does not occur. This also
follows from the fig - 3 where only $n <4$ allows \emph{flip} (for $\alpha =1)$.
Alternatively the inequality $n < 2(1+ \alpha)$  gives the condition of acceleration.
 Thus for decelerates  $V < V_{f}$ and for acceleration $V >
V_{f}$.\\
If $v_{s}$ be the velocity of sound equation \eqref{eq:eq9} gives

 \begin{equation}\label{eq:eq17a}
v_{s}^2= \left(\frac {\partial P}{ \partial \rho} \right)_S =
\frac{N \alpha}{ (1 + \alpha) \left \{1 + \left(\frac{\epsilon}{V}
\right)^N\right \} }  - \frac{n N}{n + (3N + n) \left(\frac{\epsilon}{V}\right)^N}.
 \end{equation}

The equation \eqref{eq:eq17a} offers some interesting possibilities: this gives
 $v_{s}^2 = 0$ at dust dominated universe and for large volume we get
 \begin{equation}\label{eq:eq17b}
  v_{s}^2 =  - 1 + \frac{n}{3(1+ \alpha)}.
 \end{equation}

 We shall presently see that from the thermodynamical stability conditions $n$ becomes
 negative,  leading to a phantom type of universe ~\cite{cim}. Moreover the equation  \eqref{eq:eq17b}
  gives an imaginary speed of sound for $\alpha > 0$, leading to a perturbative
  cosmology. One need not be too sceptic about it because  it favours structure formation~ \cite{fab}.

It is tempting to make some comparison with a recent work of Myong~\cite{myn}
where holographic dark energy, Chaplygin gas and tachyon model with constant
potential are briefly discussed \emph{vis-a-vis} their implications as regards
 squared acoustic speeds. This is all the more relevant because signature of the
  squared speeds is a key factor in determining the stability criteria. It is
  observed that the squared speed of tachyonic field and chaplygin gas are always
   positive while if the condition of future event horizon is assumed \emph{apriori}
    as the IR cut off but for holographic dark energy it is always negative definite.

\section*{4. Stability Criteria}

At this stage one may check the thermodynamic stability conditions of a
fluid during its evolution. For this it is necessary to find if  \emph{(i)} the
pressure reduces both for an adiabatic  and isothermal expansion~\cite{landau} $\left(
\frac{\partial P}{\partial V}\right)_{S} < 0$  $\&$ $\left(
\frac{\partial P}{\partial V}\right)_{T} < 0$
 and \emph{(ii)} as also to examine
if the thermal capacity at constant volume is positive.

 Equations  \eqref{eq:eq1} and  \eqref{eq:eq9} give

\begin{eqnarray}\label{eq:eq18}
\left( \frac{\partial P}{\partial V}\right)_{S} =  \frac{P}{3V (1+\alpha)}
\left[ \frac{3N \alpha\left( \frac{\epsilon }{V}\right)^N}{  1 +  \left(\frac
{\epsilon }{V}\right)^N } - n \right].
\end{eqnarray}

Since $P$ is always negative, $\left( \frac{\partial P}{\partial V}\right)_{S} < 0$
for $n <  \frac{3N \alpha\left( \frac{\epsilon }{V}\right)^N}{  1 +  \left(\frac
{\epsilon }{V}\right)^N } $, but this is not possible at the late stage of evolution
 because RHS of inequality is a function of volume. It will be a better option to get
  the negative value of $\left(\frac{\partial P}{\partial V}\right)_{S}$ throughout
   the evolution for both $ \alpha > 0$ and $n < 0$ simultaneously.

Now we have discussed some special cases as follows :

(i) A cursory glance at the equation  \eqref{eq:eq18} shows that $n$ and $\alpha$ can
 not be at once zero because that makes  $\left(\frac
 {\partial P}{\partial V} \right)_S =0$ leading to a severe restriction on the stability
  of the fluid. In this case the pressure becomes constant through any adiabatic change
   of volume.  Moreover $B_{0}$ here behaves like a cosmological
   constant and consequently we get a de-Sitter type of metric for late universe. When
    $\alpha = 0$  and $n \neq 0$,
    $ \left( \frac{\partial P}{\partial V} \right)_S = \frac{n}{3}
     B_0 V^{-(1+\frac{n}{3})} $, \emph{i.e.},
    for $ n < 0$, $ \left(\frac{\partial P}{\partial V} \right)_S < 0$.
     Evidently, $n \leq 0$ implies that the pressure becomes more and more
      negative with volume. This agrees well with the observational results~\cite{guo1}.

(ii) For  $n = 0$ and $ \alpha \neq 0$, the  equation \eqref{eq:eq18} becomes identical
 to an earlier work of  Santos et al~\cite{san1}.
In this case

\begin{eqnarray}\label{eq:eq18a}
 \left( \frac{\partial P}{\partial
V}\right)_{S}  =  \alpha \frac{ P}{V}
\left(\frac{\epsilon}{V} \right)^{1+\alpha} \left \{ 1+ \left(\frac{\epsilon}{V}
 \right )^{1 + \alpha} \right\}^{ - 1},
 \end{eqnarray}
 which gives  $ \left( \frac{\partial P}{\partial
V}\right)_{S}   < 0 $ for $\alpha >0$.
Again for $\alpha = 1$ and $n \neq 0$, equation  \eqref{eq:eq18} reduces to our previous
work~\cite {dp2} where

\begin{eqnarray}\label{eq:eq18b}
\left( \frac{\partial P}{\partial V}\right)_{S} =
\frac{P}{6V}\left[(6-n) \left\{1 - \frac{1}{1 +
\left(\frac{\epsilon}{V} \right)^N}\right \} -n \right]
\end{eqnarray}
This is the case of Variable Chaplygin gas(VCG).

\begin{figure}[ht]
\centering \subfigure[ This graph clearly show that $\left( \frac{\partial P}{\partial V}\right)_{S}
 < 0$ for $n < 0$ throughout the evolution.
  ]{
\includegraphics[width= 5.8 cm]{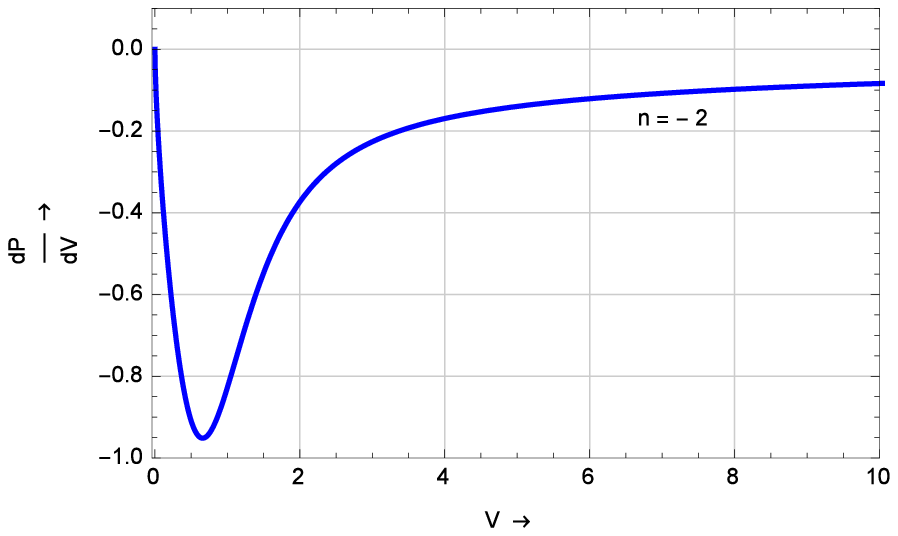}
\label{fig:subfig1} } ~~~~~\subfigure[  Here we have seen that for $ n > 0 $, VGCG does
 not stable   throughout the evolution.
 ]{
\includegraphics[width= 5.8 cm]{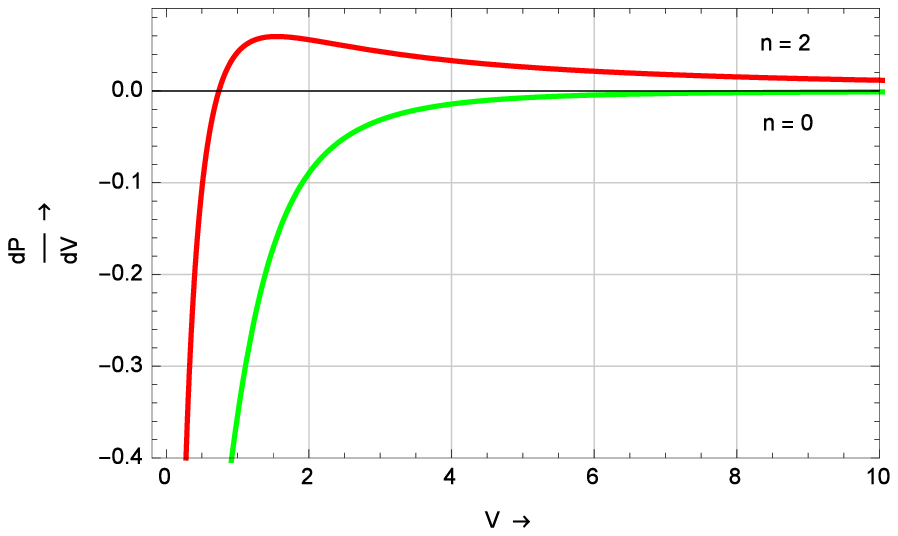}
 \label{fig:subfig2} } {

 } \label{fig:subfigureExample} \caption[Optional
caption for list of figures]{\emph{The variations of $\left( \frac{\partial P}{\partial V}
\right)_{S}$ and
   $V$   are shown}}
\end{figure}

It is clear from equation  \eqref{eq:eq18}  that for $\alpha > 0$, $n$
should be negative for $\left( \frac{\partial P}{\partial V}\right)_{S} < 0$
throughout the evolution.  It may be pointed out that this conclusion accords with
 the work of Sethi et al ~\cite{set}
  where they have obtained the best fit values of $n$ well lying between $(- 1.3, 2.6)$
   [  WMAP 1st peack + SN Ia (3 $\sigma$)] for $\alpha = 1$.
 But there is another work of  Lu~\cite{lu1} where the Union SNe Ia data
  and Sloan Digital Sky Survey(SDSS) baryon acoustic peak to  constrain
  the Variable Generalised Chaplygin Gas (VGCG) model is studied and  the
   best fit values of $n = 0.75$ for $\alpha = 1.53$ are obtained. From what
   has been discussed above we see that Lu's conclusion about the positivity of $n$
    is inadmissible when confronted with thermodynamical stability criteria.
    It also follows from fig-4.

Now we have to examine if $\left( \frac{\partial P}{\partial V}\right)_{T} \leq 0 $
 as well. We will show in the next section that for $n < 0$ this condition may also
  be satisfied.\\

\textbf{Thermal EoS:}\\

 One should also verify the positivity of thermal capacity
at constant volume $c_{V}$  where
   $c_{V}=T \left(\frac{\partial S }{\partial T} \right)_{V} =
    \left(\frac{\partial U }{\partial T} \right)_{V}=V \left(\frac{\partial
    \rho }{\partial T} \right)_{V}$. Next the temperature $T$ of the Variable
     modified Chaplygin gas as a
    function of volume $V$ and  entropy $S$ may be determined from  the relation
      $T = \left(\frac{\partial U }{\partial S}
    \right)_{V}$. As we have considered $B_{0} $ to be an absolute constant, the
      equation  \eqref{eq:eq5} gives

\begin{eqnarray}\label{eq:eq20}
T =
\frac{V^{1-N-\frac{n}{3(1+\alpha)}}}{1+\alpha}\left[\frac{B_{0}(1+\alpha)}{N}
+ V^{-N}c \right]^{-\frac{\alpha}{1+\alpha}}\left( \frac{\partial c}{\partial S} \right)_{V}   .
\end{eqnarray}

We have previously seen that $c$ may be a either universal constant or function
 of entropy $S$. If we consider $c$ to be an absolute constant $T$ becomes zero
  for all values of volume and pressure and it makes the stability criteria
   questionable. Naturally we are forced to take $c \equiv c(S)$ and temperature
    varies with expansion.  We have no \emph{a priori} knowledge of the functional
dependence of $c$. From physical considerations, however,
 we know that this function must be such as to give
positive temperature and cooling along an adiabatic expansion and
so we choose that $\left(\frac{\partial c }{\partial S}
\right)>0$.

In the literature ~\cite{san2} Santos et al have considered Jacobian identity to
 calculate the expression of $c$ for the case of Modified Chaplygin gas, where
  they assumed that $\left(
\frac{\partial P}{\partial V}\right)_{T} = 0$. But in our approach
we have considered the dimensional analysis to derive the expression of $c$
because later we will show  in this process that $\left(
\frac{\partial P}{\partial V}\right)_{T} < 0$ for $n < 0$. This is more general
approach in our sense.

Now from dimensional analysis, we observe from equation \eqref{eq:eq5} that

 \begin{equation}\label{eq:eq21}
 \left[U \right] = [ c ]^{\frac{1}{1+ \alpha}}.
\end{equation}
Using $\left[U \right] = \left[T\right]\left[S \right]$,  we can write

 \begin{equation}\label{eq:eq22}
 \left[c \right] = \left[T \right]^{1+ \alpha}\left[S \right]^{1+ \alpha}
\end{equation}
It is difficult to get an analytic solution of $c$ from equation \eqref{eq:eq22}.
 So as a trial case, we take an empirical expression of $c$ which  is a
 function of entropy only such that

\begin{equation}\label{eq:eq23}
 c = \left( \tau  \right)^{1+ \alpha} S^{1+ \alpha},
\end{equation}
where $\tau$ is a  constant having the dimension of temperature  only.
Now

\begin{equation}\label{eq:eq24}
\frac{dc}{dS} = (1+ \alpha) \left( \tau  \right)^{1+ \alpha} S^{\alpha}.
\end{equation}
Using equation \eqref{eq:eq20} and \eqref{eq:eq24}, we get the expression of temperature

\begin{subequations}\label{eq:eq25}
\begin{align}
T &=
V^{1-N-\frac{n}{3(1+\alpha)}}\left[\frac{B_{0}(1+\alpha)}{N}
+ V^{-N}c \right]^{-\frac{\alpha}{1+\alpha}}\left( \tau  \right)^{1+\alpha}S^{\alpha}
 \label{eq:eq25a} \\
&= \tau \left\{ 1- \frac{1}{1 + \left(\frac{\epsilon}{V}\right)^N }
\right\}^{\frac{\alpha}{1+\alpha}} \label{eq:eq25b}.
\end{align}
\end{subequations}

and from equation \eqref{eq:eq25},  the entropy is

\begin{equation}\label{eq:eq26}
S = \frac{\left[\frac{B_{0}(1+\alpha)}{N} V^{N} \right]^{\frac{1}{1+\alpha}}
 \left(\frac{T}{\tau^{1+\alpha}} \right)^{\frac{1}{\alpha}} }{\left \{1-\left
 (\frac{T}{\tau} \right)^{\frac{1+\alpha}{\alpha}}\right \}^{\frac{1}{1+\alpha}}}~,
\end{equation}

It follows from  equation \eqref{eq:eq26} that for positive and finite
 entropy one should have $0 < T< \tau $,  \emph{i.e.}, $\tau$ represents
  the maximum temperature.

Moreover for $T = 0$, $S = 0$  showing that the third law of thermodynamics
 is satisfied in this case.

Now using equation  \eqref{eq:eq26}
 we get the expression of thermal heat capacity as

\begin{equation}\label{eq:eq27}
c_{V}  = T\left(\frac{\partial S}{\partial T} \right)_{V} =\frac{\left
[\frac{B_{0}(1+\alpha)}{N} V^{N} \right]^{\frac{1}{1+\alpha}} \left
(\frac{T}{\tau^{1+\alpha}} \right)^{\frac{1}{\alpha}} }{ \alpha \left
 \{1-\left(\frac{T}{\tau} \right)^{\frac{1+\alpha}{\alpha}}\right \}^
 {\frac{2+ \alpha}{1+\alpha}}}~.
\end{equation}

Since $0 < T< \tau $ and $\alpha > 0$,  $c_V >0$ is always satisfied
irrespective of the value of $n$. This also ensures the positivity of $\alpha$.
It is interesting to note that when the temperature goes to zero $c_V$
 vanishes in agreement with the third law of thermodynamics.

If we now put  $\alpha = 1$, \emph{i.e.} the equation \eqref{eq:eq27}
 reduces to the expression of $c_{V}$ of our previous work~\cite{dp2}.
 Again for  $n = 0$, we get the identical expression of $c_{V}$  found
  by Santos et al~\cite{san1}.

To end the section a final remark may be in order. While
\emph{positivity} of specific heat is strongly desirable
\emph{vis-a-vis} when dealing with special relativity, in a recent
communication Luongo \emph{et al} ~\cite{luo} argued that in a FRW
type of model like the one we are discussing  a negative specific
heat at constant volume and a vanishingly small  specific heat at
constant pressure $(c_P)$ are also compatible with observational
data. In fact they have derived the most general cosmological
model which is agreeable  with the $c_{V}< 0$ and $c_{P}\sim 0$
values obtained for the specific heats of the universe and showed,
in addition, that  it also overcomes the fine-tuning and the
coincidence problems of the $\Lambda$CDM model.

We now proceed to find an expression for internal energy of  the VGCG
as a function of $V$ and $T$ with the help of equations  \eqref{eq:eq5},
 \eqref{eq:eq23} and \eqref{eq:eq26} as

\begin{equation}\label{eq:eq28}
  U= V \left \{ \frac{\frac{B_{0}(1+\alpha)}{N} V^{-\frac{n}{3}}}
  { {  1-\left(\frac{T}{\tau} \right)^{\frac{1+\alpha}{\alpha}}}}
   \right\}^{\frac{1}{1+ \alpha}}.
\end{equation}
Now using \eqref{eq:eq1}, \eqref{eq:eq2} and \eqref{eq:eq28}   the Pressure becomes

\begin{equation}\label{eq:eq29}
P  =  -  \left ( B_{0} V^{-\frac{n}{3}} \right)^{\frac{1}{1+ \alpha}}
  \left(\frac{N}{1+ \alpha} \right)^{\frac{\alpha}{1+ \alpha
  }}\left \{{  1-\left(\frac{T}{\tau} \right)^{\frac{1+\alpha}{\alpha}}}
   \right \}^{\frac{ \alpha}{1 + \alpha}}.
\end{equation}
The equation \eqref{eq:eq29} shows that for $\alpha =1 $, the expression
 for pressure is identical to our previous work~\cite{dp2} as expected.
  Further we see that for $n = 0$ the above expression goes to an earlier
   work of Santos et al~\cite{san1}. With the help of equations \eqref{eq:eq28}
    and \eqref{eq:eq29} we get the thermal EoS parameter of VGCG as

\begin{equation}\label{eq:eq29a}
 \omega =  - \frac{N}{1+ \alpha}  \left \{{  1-\left(\frac{T}{\tau}
 \right)^{\frac{1+\alpha}{\alpha}}} \right \}.
\end{equation}
 Here thermal EoS parameter is  a function of  temperature only. In the
  early era of the universe when temperature is very high, \emph{i.e.}, at $T
   \rightarrow \tau$, the  equation \eqref{eq:eq29a} gives  that $\omega \sim 0$,
   \emph{i.e.}, $P \sim 0$ representing a dust dominated universe. This is
    identical with the  equation \eqref{eq:eq12a}.  On the other hand,
     at the late stage of the universe, at very low temperature , \emph
     {i.e.}, $T \rightarrow 0$, the equation \eqref{eq:eq29a}  simplifies to
       $\omega \approx -1 + \frac{n}{3(1+ \alpha)}$. This is similar to
        equation \eqref{eq:eq13}. Thus thermodynamical state represented
        by equations \eqref{eq:eq12a} and \eqref{eq:eq13} are essentially
         same at both dust and the late stage of the universe.

Now from equation \eqref{eq:eq29} we shall  examine if $\left(\frac {\partial P}
{\partial V} \right)_{T} \leq 0 $. Thus we get

 \begin{equation}\label{eq:eq30}
  \left(\frac {\partial P}{\partial V} \right)_{T} =  \frac{n B_{0}
  V^{-(1+ \frac{n}{3})}}{3 (1+ \alpha)} \left \{ \frac{B_{0}(1+\alpha)}
  {N} V^{-\frac{n}{3}} \right\}^{\frac{1}{1+ \alpha}}\left \{{  1-\left(\frac{T}{\tau}
   \right)^{\frac{1+\alpha}{\alpha}}} \right \}^{\frac{ \alpha}{1 + \alpha}}.
\end{equation}
 which clearly shows that the value of $n$ should be negative for $\left(\frac
 {\partial P}{\partial V} \right)_{T} < 0$ throughout the evolution. It is very
  interesting to note that for $ n = 0$, $\left(\frac {\partial P}{\partial V} \right)_{T}
   = 0$. At this stage we digress to make a comparison with the works of Santos et al~
   \cite{san1, san2} referred to earlier. In their cases for GCG model they got pressure
    as a function of temperature only while for MCG they assumed the same, \emph{i.e.},
     in both the cases  $\left(\frac {\partial P}{\partial V} \right)_{T} = 0$. But in
      our case for $n < 0$,  $\left(\frac {\partial P}{\partial V} \right)_{T} <0 $
      implying that the isobaric curves of our VGCG model do not coincide with its
       isotherms in the diagram of thermodynamic states. This is definitely a
        significant improvement in our analysis.

Thus we conclude that both $ \left(\frac {\partial P}{\partial V} \right)_{S}$ and
 $ \left(\frac {\partial P}{\partial V} \right)_{T}$ are negative for $n <0$ which
  is in accordance with the stability condition of thermodynamics. One can very
   briefly check if any critical points exist in our model. We find that as in
   the case of Santos our model also does not posses any critical points.

\begin{figure}[ht]
\begin{center}
   \includegraphics[width=8cm]{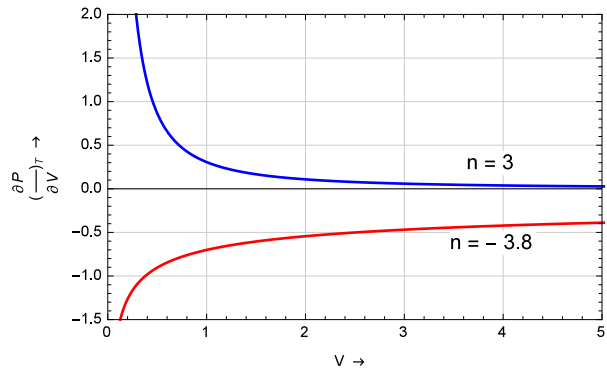}
     \caption{
  \small\emph{
  The variation of $\left(\frac {\partial P}{\partial V} \right)_{T}$ and $V$  for different
  values of $n$. We have considered here $B_{0} =1$, $\alpha = 1$ \& $c =1$.
     }\label{4}
    }
\end{center}
\end{figure}

Now from equation \eqref{eq:eq1} we get

\begin{equation}\label{eq:eq30a}
   \left(\frac{\partial P}{\partial T} \right)_{V} = \frac{\alpha B_{0}
   V^{-\frac{n}{3}}}{\rho^{\alpha +1}} \frac{\partial \rho}{\partial T}.
\end{equation}
and from equation \eqref{eq:eq28} we can write
\begin{equation}\label{eq:eq30b}
 \left(\frac{\partial U}{\partial V} \right)_{T} = \frac{N \rho}{1 + \alpha}.
\end{equation}

Now using equations \eqref{eq:eq30a} and \eqref{eq:eq30b} into  the
 wellknown relation of thermodynamics ~\cite{landau}

\begin{equation}\label{eq:eq30c}
 \left(\frac{\partial U}{\partial V} \right)_{T} = T  \left(\frac{\partial P}
 {\partial T} \right)_{V} -P,
\end{equation}
we get
\begin{equation}\label{eq:eq30d}
 \frac{\partial T}{T} = \frac{B_{0} \alpha (1 + \alpha)}{N V^{\frac{n}{3}}
  \rho^{1 + \alpha} - B_{0}} \frac{\partial \rho}{\rho}.
\end{equation}
Integrating,
\begin{equation}\label{eq:eq30e}
\beta T =  \left\{ 1- \frac{B_{0}  (1 + \alpha)}{N V^{\frac{n}{3}} \rho^{1 + \alpha} }
\right\}^{\frac{\alpha}{1 + \alpha}},
\end{equation}
where $\beta$ is an integration constant and hence we get the expression of energy
 density as

\begin{equation}\label{eq:eq30f}
  \rho=  \left \{ \frac{\frac{B_{0}(1+\alpha)}{N} V^{-\frac{n}{3}}}{
  {  1-\left(\beta T \right)^{\frac{1+\alpha}{\alpha}}}} \right\}^{\frac{1}{1+ \alpha}}.
\end{equation}
The same result can be obtained from equation \eqref{eq:eq28} if we identify
 $\beta = \frac{1}{\tau}$.
It is interesting to point out that the empirical expression of $c$ which was
taken in equation  \eqref{eq:eq23} from the consideration of dimensional analysis is justified.

As we are more concerned with the accelerating universe which is a
late stage
 phenomena where the results are similar to our previous work relating to VMCG
 model, \emph{i.e.}, at the late stage VMCG apparently reduces to VCG model.\\

We can also express the maximum temperature $\tau$ as a function of the initial
 conditions of the expansion. If we consider that the initial conditions at
   $V = V_{0}$ are $\rho = \rho_{0}$, $P = P_{0}$ and $T = T_{0}$, then we can get from \eqref{eq:eq5}

 \begin{equation}\label{eq:eq33}
    c = \left \{ \rho_{0}^{1+\alpha} - \frac{B_{0} (1 + \alpha)}{N} V_{0}^{-\frac{n}{3}}
     \right \} V_{0}^{1+\alpha}.
    \end{equation}

With the help of equations  \eqref{eq:eq6}, \eqref{eq:eq7} and \eqref{eq:eq33},
 we obtain the energy density $\rho$ and the pressure $P$ as a function of the volume $V$ as

\begin{equation}\label{eq:eq34}
   \rho = V^{- \frac{n}{3 ( 1 + \alpha)}} \rho_{0} \left [ \frac{B_{0} (1 + \alpha)}
   {N \rho_{0}^{1+ \alpha} } + \left \{ 1 - \frac{B_{0} (1 + \alpha)}{N \rho_{0}^{1+
    \alpha}} V^{-\frac{n}{3}} \right \} \left(\frac{V_{0}}{V} \right )^{1+ \alpha}
     V^{\frac{n}{3}} \right ]^{\frac{1}{1+ \alpha}},
    \end{equation}

and

\begin{equation}\label{eq:eq35}
    P = - \frac{B_{0}^{\frac{1}{1+ \alpha}}\left(\frac{B_{0}}{\rho_{0}^{1 + \alpha}}
     \right)^{\frac{\alpha}{1+ \alpha}}  V^{\frac{n}{ 3 (1 + \alpha)}}}{\left
     [ \frac{B_{0} (1 + \alpha)}{N \rho_{0}^{1+ \alpha} } + \left \{ 1 - \frac{B_{0}
      (1 + \alpha)}{N \rho_{0}^{1+ \alpha}} V^{-\frac{n}{3}} \right \} \left(\frac{V_{0}}{V}
       \right )^{1+ \alpha} V^{\frac{n}{3}} \right ]^{\frac{\alpha}{1+ \alpha}}} .
    \end{equation}

Now equations  \eqref{eq:eq29},  \eqref{eq:eq34} and  \eqref{eq:eq35} can be
written as function of the reduced parameters $\varepsilon$, $v$,  $p$, $\kappa$
and $t$ such that

\begin{eqnarray}\label{eq:eq36}\nonumber
  \varepsilon  = \frac{\rho}{\rho_{0}}, ~~~ ~~~~ v = \frac{V}{V_{0}}, ~~~p =
   \frac{P}{B_{0}^{\frac{1}{1+\alpha}}} \\
  \kappa  = \frac{B_{0} ( 1+ \alpha)}{N \rho_{0}^{1+\alpha}}, ~~~ t = \frac{T}{T_{0}}, ~~~
   \tau^{*} = \frac{\tau}{T_{0}}
\end{eqnarray}
\vspace{0.5cm}
The equations \eqref{eq:eq29},  \eqref{eq:eq34} and  \eqref{eq:eq35} can now be
 expressed in the reduced units respectively as

\begin{eqnarray}\label{eq:eq37a}
  p &=& - \left (\frac{N}{1 + \alpha} \right)^{\frac{\alpha}{1 + \alpha}}
  V^{- \frac{n}{3(1+\alpha)}}  \left \{ 1 - \left ( \frac{t}{\tau^*} \right )
  ^{\frac{1+\alpha}{\alpha}}   \right\}^{\frac{\alpha}{1+\alpha}},
  \\ \label{eq:eq37b}
    \varepsilon &=& V^{- \frac{n}{3(1+\alpha)}} \left[ \kappa + \left \{ 1 - \kappa V^
    {-\frac{n}{3}} \right \} \frac{V^{\frac{n}{3}}}{v^{1+\alpha}}  \right ]
    ^{\frac{\alpha}{1 + \alpha}},  \\ \label{eq:eq37c}
    p &=& - \frac{ \kappa^{\frac{\alpha}{1+\alpha}} \left(\frac{N V^{- \frac{n }{3 \alpha}}}
    {1 + \alpha} \right)^{\frac{\alpha}{1+\alpha}} }{\left[ \kappa +
    \left\{  1 - \kappa V^{-\frac{n}{3}} \right\} \frac{V^{\frac{n}{3}}}{v^{1+\alpha}}
     \right ]^{\frac{\alpha}{1 + \alpha}}}.
\end{eqnarray}

At $P = P_{0}$, $V = V_{0}$ and $T = T_{0}$, we have $ t =1$ and $ v = 1$ and we get
 from equations \eqref{eq:eq37a} and  \eqref{eq:eq37c},

\begin{eqnarray}\label{eq:eq38} \nonumber
    p_{0} &=& -\kappa^{\frac{\alpha}{1+\alpha}} \left( \frac{N}{1+\alpha}
    \right)^{\frac{\alpha}{1+\alpha}}V_{0}^{- \frac{ n}{3 }} , \\
        &=&   - \left (\frac{N}{1 + \alpha} \right)^{\frac{\alpha}{1 + \alpha}}
         V_{0}^{- \frac{n}{3(1+\alpha)}}  \left\{ 1 - \left ( \frac{t}{\tau^*}
         \right )^{\frac{1+\alpha}{\alpha}} \right\}^{\frac{\alpha}{1+\alpha}} ,
\end{eqnarray}

hence $\kappa$ and $\tau^*$ can be determined as follows:

\begin{equation}\label{eq:eq39}
     \kappa = V_{0}^{\frac{n}{3 }}\left\{ 1 -  \frac{1}{(\tau^*)^{\frac{1+\alpha}
     \{\alpha}}   \right\},
\end{equation}

and
\begin{equation}\label{eq:eq40}
     \tau^* = \frac{1}{\left( 1 - \kappa V_{0}^{-\frac{n}{3 }}\right)^{\frac{\alpha}{1+\alpha}}}.
\end{equation}

Interestingly, we have seen that $\tau^*$  depends on both $\kappa$, $V_{0}$ and $n$ also.
 For $n = 0$, all the above equations reduce to the equations of Santos et al ~\cite{san1,san2}.
 At the present epoch, $\kappa = \frac{B_{0} (1+ \alpha)}{N \rho_{0}^{}1+\alpha}$,
  therefore, $\rho_{0} = \left\{\frac{B_{0} (1+ \alpha)}{N \kappa} \right\}^{\frac{1}{1+\alpha}}$.
   If we consider that the temperature $\tau = 10^{32} K$ (temperature at the Planck era)
    and $T_{0} = 2.7 K$ (the temperature of the present epoch), the ratio, $\tau^{*} =
     \frac{\tau}{T_{0}} = 3.7 \times 10^{31}$. So the ratio $\kappa$ will be

\begin{equation}\label{eq:eq41}
     \kappa = V_{0}^{\frac{n}{3 }}\left \{ 1 -  \frac{1}{\left(3.7 \times  10^{31}
      \right )^{\frac{\alpha}{1+\alpha}} }  \right\} \approx V_{0}^{\frac{n}{3}},
\end{equation}
because $\alpha > 1$.
Again from \eqref{eq:eq28}, for the case of present epoch when temperature $T$ is
 very small (i.e. $T \rightarrow 0$),

\begin{equation}\label{eq:eq42}
   \rho_{0} \approx \left\{\frac{B_{0} (1+\alpha)}{N V^{\frac{n}{3}}}\right\}^
   {\frac{1}{1+\alpha}} \approx \left\{\frac{B_{0} (1+\alpha)}{N \kappa}\right\}^
   {\frac{1}{1+\alpha}},
\end{equation}

The same result can be obtained from equation \eqref{eq:eq8} for large volume.
Thus from equation \eqref{eq:eq23}, at present epoch, the energy density $\rho$
of the universe filled with VCG must be very close to $ \left\{\frac{B_{0} (1+\alpha)}
{N \kappa}\right\}^{\frac{1}{1+\alpha}}$.

\section*{5. Discussion}

 We have here studied a  very general type of exotic fluid, termed `Variable
Generalised Chaplygin gas  and discussed its cosmological
implications, mainly its thermodynamical stability.
 Although the exhaustive analysis of the latest cosmological
observations points to the existence of some form of dark energy
in the universe it is very difficult to choose among the merits of
its different forms at least from the observational results. In
fact most of the alternatives meet the energy budget. So for the
specific case of different types of Chaplygin gas we have taken
recourse to the investigation if the gas in question behaves as a
thermodynamically closed system with those values of the
parameters to meet the observational demands. In this context we
have taken the stability criteria of the gas to check its
consistency and have followed the standard prescription :
 $\left(\frac{ \partial P}{ \partial V} \right)_{S} < 0$,
 $\left(\frac{ \partial P}{ \partial V} \right)_{T} < 0$ and
  $c_{V} > 0 $. Interestingly this dictates that the new parameter, $n$
  introduced in VGCG should be negative definite.  While this conclusion accords with
  the results obtained by Sethi et al~\cite{set} where the best fit value of $n$
  lie
  between $ (- 1.3, 2.6) $ for $\alpha = 1$ but the constraint  obtained by
  Lu et al~\cite{lu1} is $n > 0$. This finding is untenable when judged from our
  results as the Lu's model becomes  thermodynamically unstable in that case.

Again our  model shows that at the dust dominated universe, the
EoS becomes $P = 0$ and at the late stage $\mathcal{ W} = -1 +
\frac{n}{3(1+\alpha)}$. As pointed earlier, from the
thermodynamical stability conditions we find that   $ n < 0$,
 which favours a  phantom model with its attendant big rip
 problem. But later the phantom like evolution is found to be compatible with
  SNe Ia observations and CMB anisotropy measurements ~\cite{cim}.

We have also studied the deceleration parameter where we calculate the
 flip volume and shows the flip is possible for $n < 2(1+\alpha)$. For
 dust dominated universe, we get $q = \frac{1}{2}$ whereas at late stage $q < 0$ for $n < 0$.

 We have  derived the expressions of the temperature as well as entropy.
  It is to be noted that at $T = 0$ both the entropy and thermal capacity of VGCG
  vanish as in conformity  with the third law of thermodynamics.
  We have  studied both the thermal and the caloric EoS which shows
   that both $0 < T < \tau$.

    From equation \eqref{eq:eq30}, it is seen that for $n <0$, $\left(\frac{ \partial P}
    { \partial V} \right)_{T} < 0$
     throughout the evolution which is also a stability condition. But we see that for
      $n = 0$, $\left(\frac{ \partial P}{ \partial V} \right)_{T} =0$. This is the case
       for Santos et al where they considered \emph{a priori} $\left(\frac{ \partial P}
       { \partial V} \right)_{T} =0$
       to calculate the expression of $c$. Actually for GCG model they got pressure as a function of
        temperature only while for MCG they assumed the same, i.e., in both the cases
        $\left(\frac{ \partial P}{ \partial V} \right)_{T} =0$. But in our case for $n <0$
         automatically $\left(\frac{ \partial P}{ \partial V} \right)_{T} < 0$ implying that
          the isobaric curves of our VGCG  model do not coincide with its isotherms in the
           diagram of the thermodynamic states. However in our model critical points are
            absent in line with the conclusion of the work of Santos et al.

    We have also expressed the equation of states in terms of reduced parameters. Again using
     wellknown thermodynamic relation \eqref{eq:eq30c} we get an expression of the energy
      density of cosmic fluid which exactly tallies with the expression \eqref{eq:eq28}.

 \vspace{0.5cm}

\textbf{Acknowledgment : } DP acknowledges  the financial support of UGC, ERO for MRP
 (No- F-PSW-165/13-14) and also Diamond Jubilee grant of Sree Chaitanya College, Habra.


\begin{thebibliography}{40}


\bibitem{reis} A. G. Reiss et al,  \emph{Astron. J.} \textbf{607} 665
(2004).

\bibitem{nop} R. A. Knop et al., \emph{Astroph. J.} \textbf{598} 102(2003).

\bibitem{aman} R. Amanullah et al,  \emph{Astrophy. J.} \textbf{716}
712 (2010).

\bibitem{spe1} D. N. Spergel  et al. (WMAP Collaboration), \emph{Astrophys. J. Suppl.} \textbf{148}  175 (2003).

\bibitem{spe2} D. N. Spergel et al. (WMAP Collaboration),  \emph{Astrophys. J. Suppl.} \textbf{170} 377 (2007).

\bibitem{neu} I. P. Neupane and H. Trowland, \emph{Int. J. Mod. Phys.} \textbf{ D19 } 364 (2010).

\bibitem{sah} V. Sahni and A. Starobinsky,\emph{ Int. J. Mod. Phys.}\textbf{ D15 } 2105  (2006).


\bibitem{dp2} D. Panigrahi, \emph{Int. J. Mod. Phys.} \textbf{D 24}, 1550030 (2015);
D. Panigrahi, Conference Proceedings : `Unified Field Mechanics: Natural Science Beyond
 the Veil of Spacetime'
edited by R. L. Amoroso, L. H. Kauffman and P. Rowlands, World Scientific, pp- 360 (2015).

\bibitem{dp4} D. Panigrahi and S. Chatterjee, \emph{JCAP} \textbf{1605} 052(2016).


\bibitem{kam}A. Kamenschik, U. Moschella and V. Pasquier, \emph{Phys. Lett.} \textbf{B511} 265 (2001).

\bibitem{ben} M. C. Bento, O. Bertolami and A. A. Sen,  \emph{Phys.
Rev. D}\textbf{66} 043507 (2002).


\bibitem{dp} D. Panigrahi and S. Chatterjee, \emph{JCAP} \textbf{10} 002(2011).

\bibitem{lu1} J. Lu, \emph{Phys. Lett.} \textbf{B680} 404
 (2009).

\bibitem{lu2} J. Lu  and L. Xu, \emph{Mod Phys. Lett.} \textbf{A250} 737 (2010).

\bibitem{set} G. Sethi et al, \emph{Int. J. Mod. Phys.} \textbf{D 15} 1089 (2006)

\bibitem{guo1} Zong-Kuan Guo  and Yuan-Zhong Zhang, \emph{Phys. Lett.} \textbf{B645} 326 (2007).

\bibitem{guo2} Zong-Kuan Guo  and Yuan-Zhong Zhang, (2005) [astro-ph/0509790 ].


\bibitem{alen} S. W. Allen et al, \emph{Mon. Not. Royl. Astro. Soc.} \textbf{353} 457 (2004).


\bibitem{san1} F. C. Santos, M L Bedran and V Soares, \emph{Phys. Lett.} \textbf{B636} 86
 (2006).

\bibitem{cim}  L. P. Cimento and R. Lazkoz, \emph{Phys. Rev. Lett.} \textbf{91} 211301 (2003);
 C. Kaconikhom, B. Gumjudpai and E. N. Saridakis,\emph{Phys. Lett.} \textbf{B695} 10 (2011).


\bibitem{fab} J. C. Fabris and J. Martin, \emph{Phys. Rev.} \textbf{D 55}  5205 (1997).

\bibitem{myn}  Y. S. Myung, \emph{ Phys lett} \textbf{B 652}  223 (2007).

 \bibitem{landau} L.  D. Landau, E.  M.  Lifschitz, Statistical Physics, third ed.,
 Course of Theoretical Physics, vol. 5, Butterworth-Heinemann, London, 1984.


\bibitem{san2} F. C. Santos, V. Soares and M. L. Bedran, \emph{Phys. Lett.} \textbf{B646} 215
 (2007).

\bibitem{luo} O. Luomgo and H. Quevedo, \emph{Gen. Rel. Grav.} \textbf{46} 1649  (2014).
\end{thebibliography}
\end{document}